\documentclass[preprint,12pt,times]{elsarticle}
\usepackage{epsfig}

\usepackage{amssymb}

\newcommand{\tbox}[1]{\mbox{\tiny #1}}

\begin{document}

\begin{frontmatter}



\title{Dynamical properties of a dissipative discontinuous map: 
A scaling investigation}

\author{R. Aguilar-S\'anchez,$^1$ Edson D. Leonel,$^2$ and J. A.
M\'endez-Berm\'udez$^3$}

\address{$^1$ Facultad de Ciencias Qu\'imicas, Benem\'erita Universidad
Aut\'onoma de Puebla, Puebla 72570, Mexico}
\address{$^2$ Departamento de F\'isica, UNESP -  Univ Estadual Paulista, Av.
24A, 1515, Bela Vista, 13506-900 Rio Claro, SP, Brazil}
\address{$^3$ Instituto de F\'isica, Benem\'erita Universidad Aut\'onoma de
Puebla, Apartado Postal J-48, Puebla 72570, Mexico}

\begin{abstract}
The effects of dissipation on the scaling properties of nonlinear
discontinuous maps are investigated by analyzing the behavior of the average
squared action $\left< I^2 \right>$ as a function of the $n$-th iteration of
the map as well as the parameters $K$ and $\gamma$, controlling nonlinearity
and dissipation, respectively. We concentrate our efforts to study the case
where the nonlinearity is large; i.e., $K\gg 1$. In this regime and for large
initial action $I_0\gg K$, we prove that dissipation produces an exponential
decay for the average action $\left< I \right>$. Also, for $I_0\cong 0$, we
describe the behavior of $\left< I^2 \right>$ using a scaling function and
analytically obtain critical exponents which are used to overlap different
curves of $\left< I^2 \right>$ onto an universal plot. We complete our study
with the analysis of the scaling properties of the deviation around the
average action $\omega$.
\end{abstract}
\end{frontmatter}

\section{Introduction}

Nonlinear maps are discrete-time dynamical systems which emerge as
local approximations to continuous-time motion, directly from a dynamical
problem, or from the successive intersections of a continuous-time trajectory
with a phase space section. Nonlinear maps have been extensively used to model 
dynamical systems in diverse knowledge areas: from maths and physics to 
finance and social sciences, passing through biology and chemistry.
However, in physics, maps are of special relevance since they can be 
directly derived from, and transformed to, Hamiltonian systems; in this case
they are named as {\it canonical maps}. Remarkably, there are 
analytical results available that describe nonlinear canonical maps; namely, 
KAM theorem and Poincar\'e--Birkhoff theorem. See for example \cite{licht}.

One of the most studied canonical maps is Chirikov's standard map (CSM), 
introduced in Ref.~\cite{C69} as a Poincar\'e surface of section of the 
kicked rotor (a prototype model for quantum chaos). CSM is an area preserving 
two-dimensional (2D) map for action and angle variables $(I,\theta)$
\begin{equation}
T_{\tbox{CSM}}:\left\{\begin{array}{ll}
I_{n+1}=I_n + K \sin(\theta_n)~,~\nonumber \\
\theta_{n+1}=\theta_n + I_{n+1}~,~\quad \mbox{mod}(2\pi) \\
\end{array}
\right.~.
\label{CSM}
\end{equation}
CSM, which is identified as a {\it continuous} map due to the sine function, 
describes the situation when nonlinear resonances are equidistant in phase
space which corresponds to a local description of dynamical chaos
\cite{licht}. Due to this property various dynamical systems and maps can be
locally reduced to map (\ref{CSM}). Thus, CSM describes the universal and
generic behavior of nearly-integrable Hamiltonian systems with two degrees of
freedom having a divided phase space composed of stochastic motion bounded by
invariant tori (known as KAM scenario) \cite{licht}.

CSM develops two dynamical regimes separated by the critical parameter 
$K_{\tbox{C}}\approx 0.971635\ldots$ \cite{licht,C69,C79,G79,M83,MMP84,MP85}. 
When $K<K_{\tbox{C}}$, known as regime of weak nonlinearity, the motion is mainly 
regular with regions of stocasticity where the action $I$ is bounded by KAM 
surfaces. At $K=K_{\tbox{C}}$, the last KAM curve is destroyed 
and the transition to global stocasticity takes place. 
Then, for $K>K_{\tbox{C}}$, regime of strong nonlinearity, $I$ becomes unbounded 
and increases diffusively.

In particular, in Ref.~\cite{LS07} a scaling analysis of CSM was performed 
by studying the average value of the squared action $\left< I^2 \right>$ as a 
function of $K$ and the $n$-th iteration of the map. There, the following
scaling law was reported:
\begin{eqnarray} 
\left< I^2 \right>\propto n^\alpha K^\beta \ ;
\label{scaling} 
\end{eqnarray}
where $\alpha\approx 2$ for $K\ll K_{\tbox{C}}$ and small\footnote{Note that
in the regime of weak nonlinearity, $K<K_{\tbox{C}}$, the action $I$ is bounded;
however, the scaling of Eq.~(\ref{scaling}) is observed before $\left< I^2 \right>$
saturates.} $n$ while $\alpha\approx 1$ for $K\gg K_{\tbox{C}}$ and large $n$, 
with $\beta\approx 2$ in both cases.\footnote{It is also relevant to mention that 
the scaling of Eq.~(\ref{scaling}) is valid regardless the value of $I_0$:
$I_0<K$ (even $I_0=0$) or $I_0>K$. However, for $I_0>K$ the behavior of 
$\left< I^2 \right>$ versus $n$ becomes richer; we refer the reader to 
Ref.~\cite{LS07} for details.}
The scaling of Eq.~(\ref{scaling}) turned out to be more general 
and has also been observed for other dynamical systems that can be represented
locally (but not exclusively) by the standard map such as the Fermi-Ulam model
\cite{ulam1,ulam3,osiel1,osiel2}, time-dependent potential wells
\cite{well1}, waveguide billiards \cite{waveguide}; among other nonlinear
systems \cite{other1,other2}.

In this paper we consider as dynamical model a dissipative version of the
discontinuous CSM. We will seek to understand and describe the behavior of 
the squared action as a function of the control parameters of the map as well 
as the time. To this end we will make use of analytical treatments supported 
by extensive numerical simulations. For the regime of large
nonlinearity in the presence of dissipation, when the dynamics starts with
small initial action, the curves we obtain for the averaged square action 
show a clear growth that is stopped when the curves reach a regime of saturation.
The saturation is produced by the presence of the dissipation. Indeed it is
a consequence of the fact that the determinant of the Jacobian matrix is smaller
than one. Such property leads to the existence of attractors in the phase
space. Since such attractors are far away from infinity, the trajectories can
not diffuse without limit, in contrast to the nondissipative case. So
this is a clear transition from unlimited to limited diffusion in the action. 
For the best knowledge of the authors, we found analytically the exponents
describing such a dynamical regime for the first time. Our results are totally 
supported by large scale numerical simulations. 
In addition, when the initial action is large enough, say few orders of magnitude larger 
than the nonlinearity, we prove analytically that the decay of the action is
exponential. Also, numerical simulations are used to validate the analytical findings, 
giving total support for our results. 
The procedures we used in this paper are general and may be extended to different 
types of nonlinear models, particularly to higher dimensional mappings.

\section{Model and numerical procedure}
\label{sec2}

Indeed, even though 
CSM describes the universal behavior of area-preserving continuous maps,
other class of Hamiltonian dynamical systems are represented by the so
called {\it discontinuous} map \cite{B98}
\begin{equation}
T_{\tbox{DCSM}}:\left\{\begin{array}{ll}
I_{n+1}=I_n + K \sin(\theta_n)f(\theta_n)~,~\nonumber \\
\theta_{n+1}=\theta_n + \tau I_{n+1}~,~\quad \mbox{mod}(2\pi) \\
\end{array}
\right.~,
\label{DM}
\end{equation}
where 
\begin{equation}
f(\theta_n)=\mbox{sgn}[\cos(\theta_n)] \ .
\label{f}
\end{equation}
There are several physical
systems that can be described by discontinuous maps,
including 2D billiard models like the stadium billiard
\cite{stadium1,stadium2} and polygonal billiards \cite{poly1,poly2}. The
origin of the discontinuity in map (\ref{DM}) are the sudden translations of
the action under the system dynamics.

In the same way as CSM, mapping (\ref{DM}) is known to have two different
dynamical regimes delimited by the critical parameter $K_{\tbox{C}}=1/\tau$ 
\cite{B98}; nevertheless, both of them are diffusive. 
The regimes $K<K_{\tbox{C}}$ and $K>K_{\tbox{C}}$ are known
as slow diffusion and quasilinear diffusion regimes, respectively. 
On the one hand, the main difference between CSM and map 
(\ref{DM}) is that for $K<K_{\tbox{C}}$
the later does not show regular behavior. In fact, due to the discontinuity
of $f(\theta)$, KAM theorem is not satisfied and map
(\ref{DM}) does not develop the KAM scenario. Since for any $K\ne 0$ the
dynamics of map (\ref{DM}) is diffusive, a single trajectory can explore the
entire phase space. However, in the slow diffusion regime the dynamics is far
from being stochastic due to the sticking of trajectories along cantori
(fragments of KAM invariant tori). On the other hand, for $K>K_{\tbox{C}}$ map 
(\ref{DM}) shows diffusion similar to that of CSM.

Following the same general procedure reported in Ref.~\cite{LS07}, in 
Ref.~\cite{MA12} it was found that the scaling of $\left< I^2 \right>$ for 
discontinuous maps when $K\ll K_{\tbox{C}}$ and $K\gg K_{\tbox{C}}$ obeys the same
scaling laws, in the appropriate limits, than CSM in the regimes of weak and
strong  nonlinearity, respectively. 
However, due to absence of KAM tori, it was observed that $\left< I^2 \right>$ 
is described as a power law of the type $\left< I^2 \right>\propto n K^{\beta}$;
which applies in both regimes (for large enough $n$) with
$\beta\approx 5/2$ when $K\ll K_{\tbox{C}}$ and $\beta\approx 2$ for $K\gg K_{\tbox{C}}$.

Since the purpose of this work is to study the effects of dissipation on 
the scaling properties of a discontinuous map, to consider a dissipative
dynamics we propose the following mapping
\begin{equation}
T_{\tbox{DDCSM}}:\left\{\begin{array}{ll}
I_{n+1}=(1-\gamma)I_n + K \sin(\theta_n)f(\theta_n)~,~\nonumber \\
\theta_{n+1}=\theta_n + \tau I_{n+1}~,~\quad \mbox{mod}(2\pi) \\
\end{array}
\right.~,
\label{DDM}
\end{equation}
where $\gamma\in[0,1]$ is the dissipation parameter and the function
$f(\theta_n)$ is given by Eq.~(\ref{f}). If
$\gamma=0$ in (\ref{DDM}) the Hamiltonian area-preserving discontinuous map of
Eq.~(\ref{DM}) is recovered. We shall consider $0<\gamma<1$. Given the
determinant of the Jacobian matrix is $1-\gamma$, the system is area
preserving only when $\gamma=0$. For mapping (\ref{DDM}) we study some of the
properties for the scaling of the average value of the squared action variable
$\left< I^2 \right>$ and of the deviation around the average action $\omega$, 
as a function of $n$, $K$, $I_0$, and $\gamma$.

We compute $\left< I^2 \right>$ for map (\ref{DDM}) following two steps: First we
calculate the average squared action over the orbit associated with the
initial condition $j$ as
\begin{equation}
\left< I^2_{n,j} \right> = \frac{1}{n+1} \sum^n_{i=0} I^2_{i,j} \ ,
\label{I2a}
\end{equation}  
where $i$ refers to the $i$-th iteration of the map. Then, the average value
of $I^2$ is defined as the average over $M$ independent realizations of the
map (by randomly choosing values of $\theta_0$):
\begin{equation}
\left< I^2 \right>(n,K,I_0,\gamma) = 
\frac{1}{M} \sum^M_{j=1} \left< I^2_{n,j} \right> \ .
\end{equation}
Then, we define the average standard deviation of $I$ as
\begin{equation}
\omega(n,K,I_0,\gamma) = \frac{1}{M} \sum^M_{j=1} 
\sqrt{ \left< I_{n,j}^2 \right> - \left< I_{n,j} \right>^2 } \ ,
\label{w}
\end{equation}
where, in analogy with Eq.~(\ref{I2a}), $\left< I_{n,j} \right> = (n+1)^{-1}
\sum^n_{i=0} I_{i,j}$. 

In the following, without lose of generality, we set $\tau=1$. 
In our simulations, for each combination of parameters $(K,I_0,\gamma)$, we 
consider an ensemble of $1000$ different initial random phases uniformly 
distributed in the interval $0<\theta_0<2\pi$. A larger ensemble leads to
qualitatively the same results.

\section{Results. Case $K\gg 1$}
\label{sec3}

We divide our investigation in two parts, both considering $K\gg 1$. 
The first is devoted to $I_0\cong 0$ while the second takes into account $I_0\gg 1$.

\subsection{$I_0\cong 0 \ll K$}

\begin{figure}
\begin{center}
\includegraphics[width=8cm]{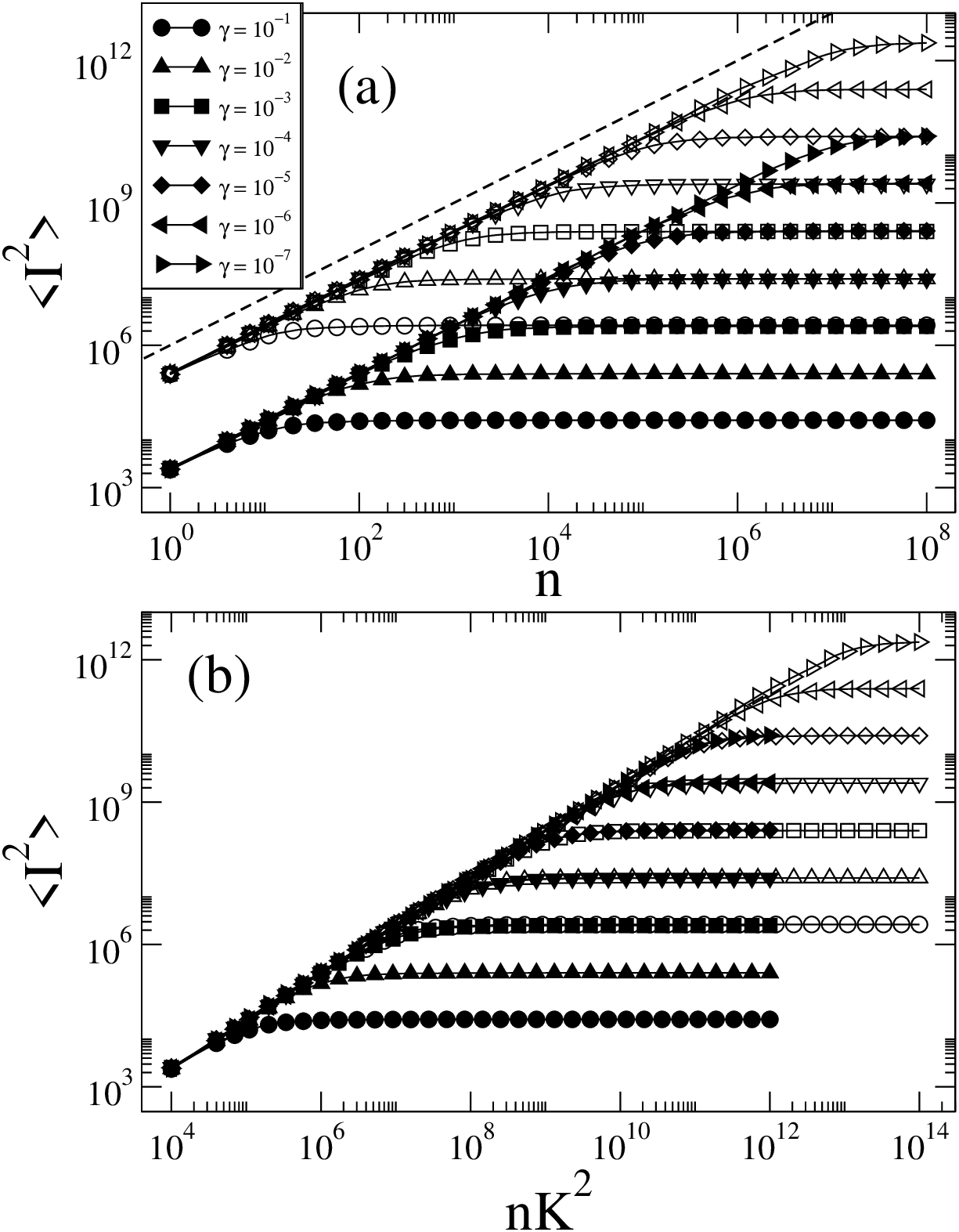}
\end{center}
\caption{$\left< I^2 \right>$ as a function of (a) $n$ and (b) $nK^2$ 
for the map of Eq.~(\ref{DDM}) when $K\gg1$ and $I_0\ll K$. Full symbols 
(open symbols) correspond to $K=10^2$ ($K=10^3$). $I_0=K/100$ was 
used. 
The dashed line in (a) proportional to $n$ is plotted to guide the eye.
In (b), after the transformation $n\rightarrow nK^2$, all curves grow 
together.}
\label{Fig1}
\end{figure}

Because of the symmetry of the phase space from both positive ($I>0$) and 
negative ($I<0$) sides, the average $\left< I \right>$ is not the most convenient 
variable to look at when $I_0\cong 0$. Instead, we consider $\left< I^2 \right>$.
Figure \ref{Fig1}(a) shows plots of $\left< I^2 \right>$ as a function of $n$ for 
several values of $\gamma$ and two different values of $K$: $K=10^2$ and $K=10^3$.
For fixed $K$, we observe that the curves $\left< I^2 \right>$ vs.~$n$ show
two different regimes: a growth regime, for small iteration values, and a  
saturation regime marked by a constant plateau $I^2_{\tbox{sat}}$, for
large enough $n$. The transition from growth to saturation is
characterized by the crossover iteration number $n_x^{(1)}$. We notice that
different values of $K$ make the curves of $\left< I^2 \right>$ to grow parallel 
to each other with the same slope, leading us to believe that $n$ is not a
good scaling variable. In fact, by applying the transformation 
$n\rightarrow nK^2$ the growth part of all curves $\left< I^2 \right>$ vs.~$n$ 
coalesce, as shown in Fig.~\ref{Fig1}(b).

Based on the behavior shown in Fig.~\ref{Fig1} we 
propose the following scaling hypotheses for $\left< I^2 \right>$:
\begin{enumerate}
\item[(i)]{$\left< I^2 \right>(nK^2,K,\gamma)\propto (nK^2)^{\beta}$, for $n\ll n_x^{(1)}$,
where $\beta$ is an accelerating exponent;
}
\item[(ii)]{$I^2_{\tbox{sat}}\propto K^{\alpha_1}\gamma^{\alpha_2}$, for $n\gg n_x^{(1)}$,
where both $\alpha_1$ and $\alpha_2$ are scaling exponents;
}
\item[(iii)]{$n_x^{(1)}K^2\propto K^{z_1}\gamma^{z_2}$, with $z_1$ and $z_2$
representing dynamical exponents.
}
\end{enumerate}

\begin{figure}
\begin{center}
\includegraphics[width=12cm]{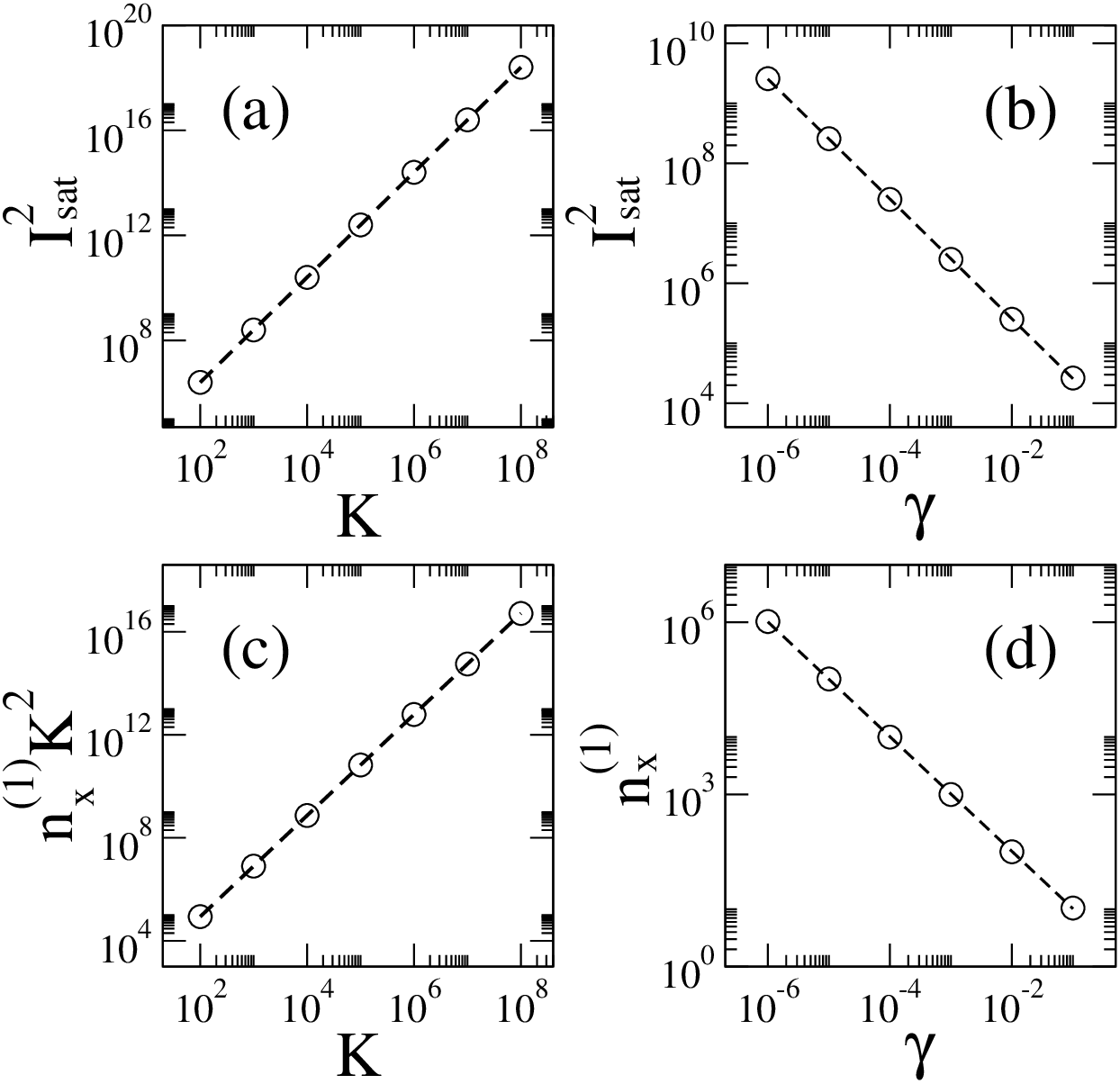}
\end{center}
\caption{$I^2_{\tbox{sat}}$ as a function of (a) $K$ and (b) $\gamma$. 
(c) $n_x^{(1)}K^2$ as a function of $K$.
(d) $n_x^{(1)}$ as a function of $\gamma$.
In (a) and (c) $\gamma$ was set to $10^{-3}$.
In (b) and (d) $K$ was set to $10^2$.
In all cases $I_0=K/100$ was used.
Power law fittings to the data demonstrate the scalings 
(a) $I^2_{\tbox{sat}}\propto K^{\alpha_1}$ with $\alpha_1=2$, 
(b) $I^2_{\tbox{sat}}\propto\gamma^{\alpha_2}$ with $\alpha_2=-1$,
(c) $n_x^{(1)}K^2\propto K^{z_1}$ with $z_1=2$, and 
(d) $n_x^{(1)}\gamma^{z_2}$ with $z_2=-1$; 
see dashed lines in the corresponding panels.}
\label{Fig2}
\end{figure}

As a consequence of these scaling hypotheses, we can describe the 
behavior of $\left< I^2 \right>$ using a homogeneous function of the type
\begin{equation}
\left< I^2 \right>(nK^2,K,\gamma)=\ell \left< I^2 \right>(\ell^anK^2,\ell^bK,\ell^c\gamma)~,
\label{homogeny}
\end{equation}
where $\ell$ is a scaling factor and $a$, $b$, and $c$ are scaling
exponents. In fact, since $\ell$ is an arbitrary scaling factor, we can make the 
following independent choices:
\begin{itemize}
\item
$\ell^anK^2=1$. 

Then, substituting $\ell=(nK^2)^{-1/a}$ into (\ref{homogeny}) leads to
\begin{equation}
\left< I^2 \right>(nK^2,K,\gamma)=(nK^2)^{-1/a}\left< I^2 \right>_1((nK^2)^{-b/a}K,(nK^2)^{-c/a}
\gamma)~,
\label{Esc_1}
\end{equation}
where $\left< I^2 \right>_1((nK^2)^{-b/a}K,(nK^2)^{-c/a}
\gamma)=\left< I^2 \right>(1,(nK^2)^{-b/a}K,(nK^2)^{-c/a}
\gamma)$ is assumed to be constant for $n\ll n_x^{(1)}$. 
By comparing Eq.~(\ref{Esc_1}) with the first scaling hypotheses, we end up 
with $\beta=-1/a$.

The accelerating exponent $\beta$ can be obtained by performing power-law 
fittings to the curves $\left< I^2 \right>$ vs.~$n$ for $n\ll n_x^{(1)}$.
Indeed, we found $\beta=1$ from several different simulations, 
see the dashed line in Fig.~\ref{Fig1}(a). 
\item
$\ell^bK=1$. 

After substitution of $\ell=K^{-1/b}$ into Eq.~(\ref{homogeny}) we get
\begin{equation}
\left< I^2 \right>(nK^2,K,\gamma)=K^{-1/b}\left< I^2 \right>_2(K^{-a/b}nK^2,K^{-c/b}\gamma)~,
\label{Esc_2}
\end{equation}
where
$\left< I^2 \right>_2(K^{-a/b}nK^2,K^{-c/b}\gamma)=\left< I^2 \right>(K^{-a/b}nK^2,1,K^{-c/b}\gamma)$ is assumed to be constant for $n\gg n_x^{(1)}$. By comparing Eq.~(\ref{Esc_2}) with
the second scaling hypotheses we obtain that $\alpha_1=-1/b$. 

In Fig.~\ref{Fig2}(a) we plot $I^2_{\tbox{sat}}$ vs.~$K$ for fixed $\gamma$. 
There, the power law $I^2_{\tbox{sat}}\propto K^{\alpha_1}$ is clearly observed. 
A power-law fitting to the data provides $\alpha_1=2$.
\item
$\ell^c\gamma=1$.

By substituting $\ell=\gamma^{-1/c}$ into Eq.~(\ref{homogeny}) we obtain
\begin{equation}
\left< I^2 \right>(nK^2,K,\gamma)=\gamma^{-1/c}\left< I^2 \right>_3(\gamma^{-a/c}nK^2,\gamma^{-b/c}K)~,
\label{Esc_3}
\end{equation}
where
$\left< I^2 \right>_3(\gamma^{-a/c}nK^2,\gamma^{-b/c}K)=\left< I^2 \right>(\gamma^{-a/c}nK^2,\gamma^{-b/c}
K,1)$ is supposed to be constant for $n\gg n_x^{(1)}$. Thus, from the comparison 
of Eq.~(\ref{Esc_3}) with the second scaling hypotheses, we get 
$\alpha_2=-1/c$. 

Then, in Fig.~\ref{Fig2}(b) we present $I^2_{\tbox{sat}}$ vs.~$\gamma$ for fixed $K$. 
From this figure the power law $I^2_{\tbox{sat}}\propto \gamma^{\alpha_2}$ is evident. 
Here, we obtain $\alpha_2=-1$ from the power-law fitting of the data.

\end{itemize}

To obtain the dynamical exponents $z_1$ and $z_2$, we can use the three 
different expressions for $\ell$
we obtained above: (1) $\ell=(nK^2)^\beta$, (2) $\ell=K^{\alpha_1}$, and 
(3) $\ell=\gamma^{\alpha_2}$. 
Indeed by comparing (1) with (2) we get $(nK^2)^{\beta}\propto K^{\alpha_1}$, for
fixed $\gamma$. Therefore, $nK^2\propto K^{\alpha_1/\beta}$, leading to the scaling law
\begin{equation}
z_1={{\alpha_1}\over{\beta}}~.
\label{Eq_z1}
\end{equation}
Comparing now (2) and (3) for a fixed $K$, we obtain that
$(nK^2)^{\beta}\propto \gamma^{\alpha_2}$, producing the scaling law
\begin{equation}
z_2={{\alpha_2}\over{\beta}}~.
\label{Eq_z2}
\end{equation}
Therefore, since we already know the values of $\alpha_{1,2}$ and $\beta$ we can compute 
the values of the dynamical exponents as: $z_1=2$ and $z_2=-1$. 
Numerical simulations confirm these values for $z_{1,2}$, as shown in Figs.~\ref{Fig2}(c) and \ref{Fig2}(d), where we plot $n_x^{(1)}K^2$ vs.~$K$ and $n_x^{(1)}$ vs.~$\gamma$, respectively.

\begin{figure}
\begin{center}
\includegraphics[width=8cm]{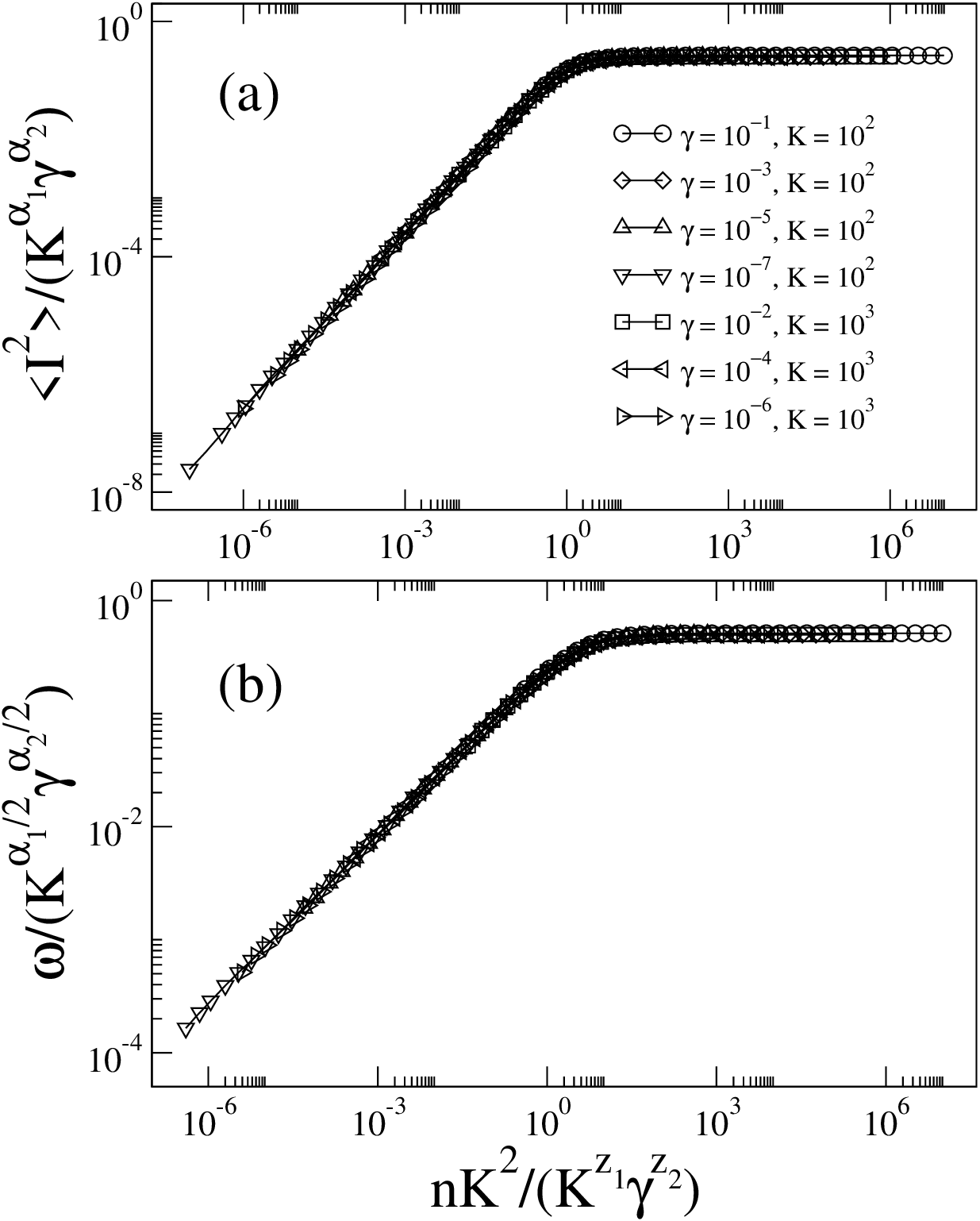}
\end{center}
\caption{(a) $\left< I^2 \right>$ and (b) $\omega$ as a function of 
$nK^2$ for the map of Eq.~(\ref{DDM}) when $K\gg1$ and $I_0\ll K$. The curves are 
properly scaled on both axis to fall on top of a universal plot.
Several values of $K$ and $\gamma$ were used with $I_0=K/100$.}
\label{Fig4}
\end{figure}

The exponents defined above can be used on the transformations 
$\left< I^2 \right>\rightarrow \left< I^2 \right>/(K^{\alpha_1}\gamma^{\alpha_2})$ 
and $n\rightarrow nK^2/(K^{z_1}\gamma^{z_2})$ to overlap the curves  
$\left< I^2 \right>$ vs.~$n$ onto a single and therefore universal curve.
This is shown in Fig.~\ref{Fig4}(a), where we clearly observe that 
all curves 
$\left< I^2 \right>/(K^{\alpha_1}\gamma^{\alpha_2})$ vs.~$nK^2/(K^{z_1}\gamma^{z_2})$
fall one on top of the other for several combinations of $K$ and $\gamma$.

We found a similar behavior for the observable $\omega$, which is the deviation 
around the average action. Since $\omega^2\propto \left< I^2 \right>$,
we could conclude that
\begin{enumerate}
\item[(i)]{$\omega\propto (nK^2)^{\beta/2}$, for $n\ll n_x^{(1)}$;
}
\item[(ii)]{$\omega_{\tbox{sat}}\propto K^{\alpha_1/2}\gamma^{\alpha_2/2}$, for $n\gg
n_x^{(1)}$;
}
\item[(iii)]{$n_x^{(\omega)}\propto K^{z_1}\gamma^{z_2}$. 
}
\end{enumerate}
Here, the exponents $\beta$, $\alpha_{1,2}$, and $z_{1,2}$ are the same exponents we 
defined for the scaling of $\left< I^2 \right>$.

The validation of the scaling hypotheses for $\omega$ is shown in Fig.~\ref{Fig4}(b)
where we show the collapse of all curves 
$\omega/(K^{\alpha_1}\gamma^{\alpha_2})$ vs.~$nK^2/(K^{z_1}\gamma^{z_2})$
onto a single universal curve for several combinations of $K$ and $\gamma$.

\subsubsection{Analytical finding of the critical exponents $\beta$,
$\alpha_1$, and $\alpha_2$}

In this subsection we give analytical arguments to support the values we
obtained numerically for the critical exponents $\beta$, $\alpha_1$, and
$\alpha_2$.

On the one hand, we know that $\left< I_n \right>=0$ due to the symmetry of
the phase space. Hence, the dispersion of $I$, indeed $\omega$, must behave
as $\omega \propto \sqrt{n}$, therefore $\left< I^2 \right>\propto n$ for $n\ll n_x^{(1)}$.
Using the transformation $n\rightarrow nK^2$ as made previously we obtain
$\left< I^2 \right>\propto nK^2$. A comparison with the scaling hypothesis (i) leads to $\beta=1$.

On the other hand, at equilibrium, i.e. at saturation, we must have $\left<
I^2_{n+1} \right> = \left< I^2_n \right> = I^2_{\tbox{sat}}$, hence
\begin{eqnarray}
\label{I2K2}
\left< I^2_{n+1} \right> 
& = & \left< \left[(1-\gamma)I_n +
K\sin(\theta_n)f(\theta_n)\right]^2 \right> \nonumber \\
& = & (1-\gamma)^2 \left< I_n^2 \right> + 
K^2 \left< \sin^2(\theta_n)f^2(\theta_n) \right> \nonumber \\
&& + \ 2(1-\gamma)K \left< I_n \right> \left< \sin(\theta_n)f(\theta_n)
\right> \nonumber \\
& = & (1-\gamma)^2 \left< I_n ^2 \right> + 
K^2 \left< \sin^2(\theta_n)f^2(\theta_n) \right> = \left< I^2_n \right> =
I^2_{\tbox{sat}} \ . \nonumber
\end{eqnarray}
The term $2(1-\gamma)K \left< I_n \right> \left< \sin(\theta_n)f(\theta_n)
\right>$ above was eliminated because both $\left< I_n \right>=0$ (given the
symmetry of the phase space) and $\left< \sin(\theta_n)f(\theta_n) \right>=0$.
Therefore, $(1-2\gamma+\gamma^2) I^2_{\tbox{sat}} + K^2 \left<
\sin^2(\theta_n)f^2(\theta_n) \right> = I^2_{\tbox{sat}}$
or
$\gamma(\gamma-2) I^2_{\tbox{sat}} = -K^2 \left< \sin^2(\theta_n)f^2(\theta_n) \right>$,
which leads to
\begin{equation}
I^2_{\tbox{sat}} = \frac{\left< \sin^2(\theta_n)f^2(\theta_n)
\right>}{2-\gamma} K^2 \gamma^{-1} 
= CK^2 \gamma^{-1} \ ,
\label{Isat}
\end{equation}
where we have set $\left< \sin^2(\theta_n)f^2(\theta_n)\right>/(2-\gamma)=C$ 
as a constant. From Eq.~(\ref{Isat}) we conclude
that $I^2_{\tbox{sat}} \propto K^2\gamma^{-1}$, meaning that $\alpha_1=2$ and
$\alpha_2=-1$, see scaling (ii). We just recall that these values for $\alpha_1$ 
and $\alpha_2$ have been already numerically validated in Figs.~\ref{Fig2}(a) and 
\ref{Fig2}(b), respectively.

\subsection{$I_0\gg K$: Exponential decay of action}

Let us now consider the case of large initial action: $I_0\gg K$. For large
$I$, the variable $\theta_{n+1}$ becomes strongly uncorrelated with
$\theta_n$. It therefore makes the function $\sin(\theta_n)
\mbox{sgn}[\cos(\theta_n)]$ to fluctuate very fast. Then, for sufficiently
small $K$, as compared to $I_0$, we can write $I_n\approx (1-\gamma)^nI_0$
that for $n\gg 1$ gets the form
\begin{equation}
I_n \approx I_0 \exp(-\gamma n) \ ,
\label{In}
\end{equation}
which predicts the exponential decay of the action with a decay 
rate equal to the dissipation parameter $\gamma$.

\begin{figure}
\begin{center}
\includegraphics[width=8cm]{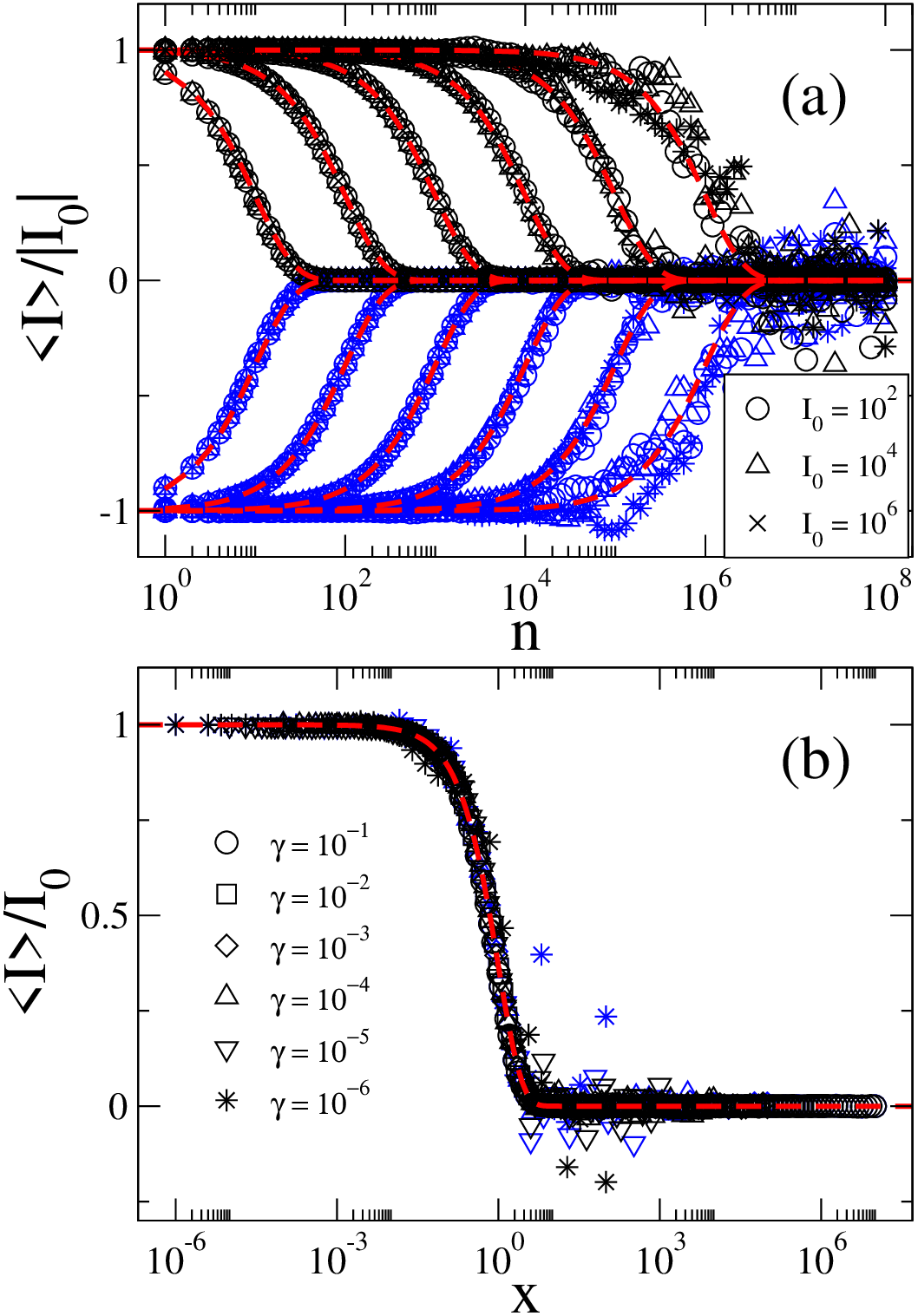}
\end{center}
\caption{(a) $\left\langle I \right\rangle/\mid I_0 \mid$ as a
function of $n$ and (b) $\left\langle I \right\rangle/I_0$ as a function of
$x$, where $x=\gamma n$, for the map of Eq.~(\ref{DDM}) when $K\gg1$ and
$I_0\gg K$. $K=I_0/100$ was used. Black (blue) symbols correspond to $I_0>0$
($I_0<0$). In (a) the groups of symbols correspond to $\gamma=10^{-1}$,
$10^{-2}$, $10^{-3}$, $10^{-4}$, $10^{-5}$, and $10^{-6}$, from left to
right. In (b) all symbols correspond to $I_0=10^3$. Red dashed lines in (a)
[(b)] are $\pm \exp(-\gamma n)$ [$\exp(-x)$].}
\label{Fig6}
\end{figure}

To verify Eq.~(\ref{In}) in Fig.~\ref{Fig6}(a) we plot 
$\left\langle I \right\rangle/\mid I_0 \mid$ as a function of $n$ for several 
combinations of $I_0$ and $\gamma$. We are including Eq.~(\ref{In}) as red dashed 
lines. It is clear from this figure that 
Eq.~(\ref{In}) reproduces well the numerically obtained data, mainly for {\it 
large} $\gamma$: that is, when $\gamma\to 0$ the damping is weak and do not 
produce a clean exponential decay of $I$; see for example the curves for 
$\gamma=10^{-6}$ in Fig.~\ref{Fig6}(a) (right-most curves) which show a decay 
that is not well described by Eq.~(\ref{In}).
Moreover, due to the excellent correspondence between theory and
numerics for {\it large} enough $\gamma$, in Fig.~\ref{Fig6}(b)
we show the overlap of the curves $\left\langle I \right\rangle/I_0$ 
when plotted as a function of $x\equiv\gamma n$. 

\begin{figure}
\begin{center}
\includegraphics[width=8cm]{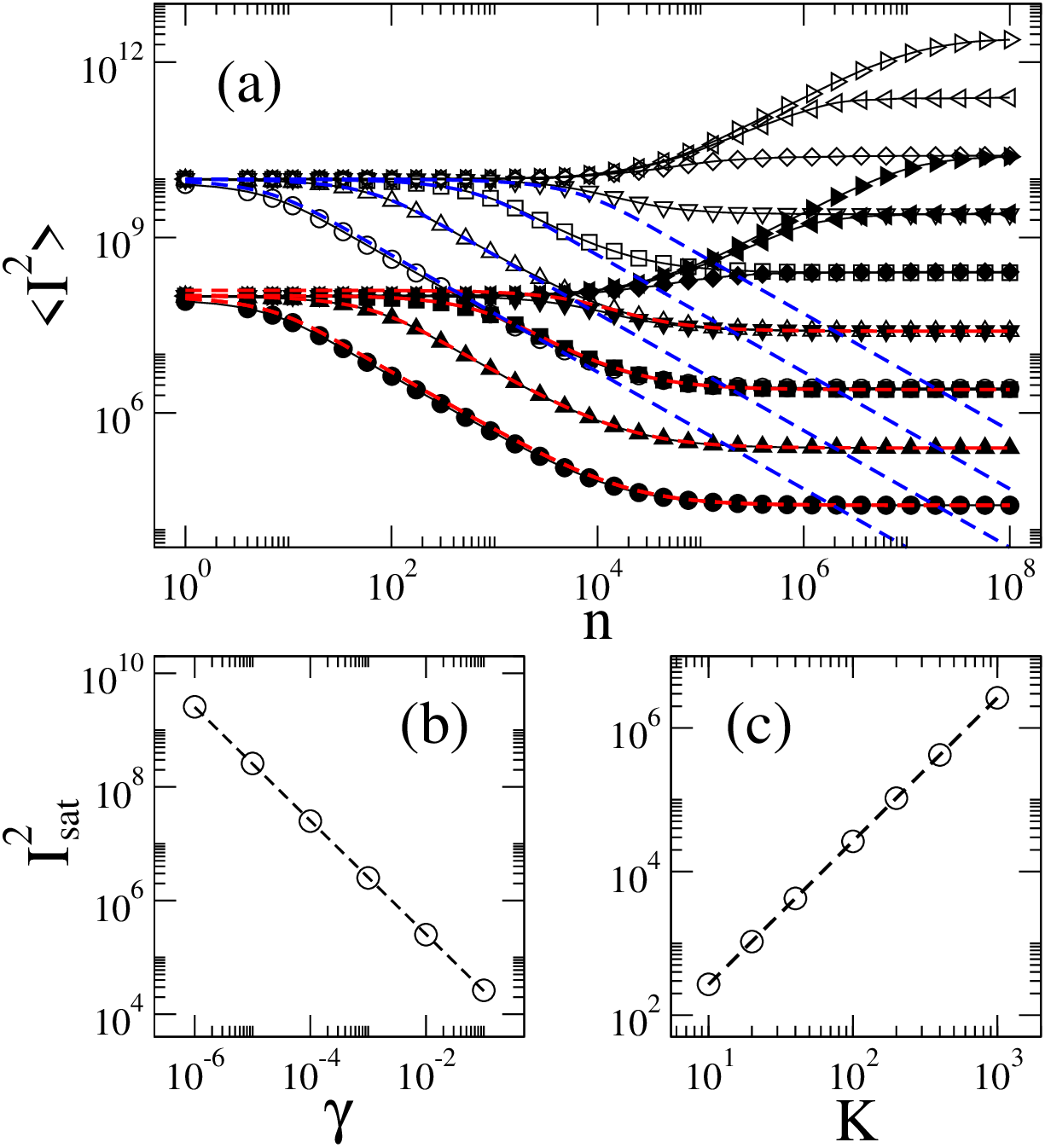}
\end{center}
\caption{(a) $\left< I^2 \right>$ as a function of $n$ for the
map of Eq.~(\ref{DDM}) when $K\gg1$ and $I_0\gg K$. Full (open) symbols
correspond to $K=10^2$ ($K=10^3$). $I_0=100K$ was used. The curves in each
set, full symbols and open symbols, correspond to $\gamma=10^{-1}$, $10^{-2}$,
$10^{-3}$, $10^{-4}$, $10^{-5}$, $10^{-6}$, and $10^{-7}$, from bottom to top.
Blue [Red] dashed lines are Eq.~(\ref{I2}) [Eq.~(\ref{I2Isat})]. (b) [(c)]
$I^2_{\tbox{sat}}$ as a function of $\gamma$ [$K$] for $K=10^2$ and $I_0=10^4$
[$\gamma=10^{-1}$ and $I_0=10^4$]. The power-law fittings in (b-c), see dashed lines, 
demonstrate the scalings $I^2_{\tbox{sat}}\propto\gamma^{\alpha_2}$ and 
$I^2_{\tbox{sat}}\propto K^{\alpha_1}$ with $\alpha_2=-1$ and $\alpha_1=2$, 
respectively.}
\label{Fig7}
\end{figure}

Even though Eq.~(\ref{In}) provides a suitable global description of 
$\left\langle I \right\rangle$ vs.~$n$, we stress that the oscillations 
of $\left\langle I \right\rangle$ around zero for large $n$ do not disappear 
when $n\to \infty$; mainly for $\gamma\to 0$ where the oscillations are
in fact quite large. This fact should manifest itself in the behavior
of $\left< I^2 \right>$ and $\omega$ making them grow (instead of decrease) 
in an intermediate interval of $n$, as will be shown below. This is indeed
expected since the approach developed above is valid for $I\gg K$ only, 
which may not be satisfied when $n$ is large and $\gamma$ small.

Now, let us explore the behavior of $\left< I^2 \right>$. In
Fig.~\ref{Fig7}(a) we plot $\left< I^2 \right>$ as a function of 
$n$ for several values of $\gamma$. We have used two values of $I_0$: $I_0=10^4$ 
and $I_0=10^5$. According to this figure, the behavior of $\left< I^2 \right>$ as
a function of $n$ is as follows. During the first iteration steps, since $K$
is small as compared to $I_0$, $\left< I^2 \right>$ does not change significantly as a
function of $n$; so $\left< I^2 \right>$ remains approximately equal to
$I_0^2$ up to a crossover iteration number $n_x$. Then, when $n>n_x$, $\left<
I^2 \right>$ shows two different behaviors. On the one hand, for small
$\gamma$, let us say for $\gamma<\gamma_{\tbox{C}}$, $\left< I^2 \right>$ increases as
a function of $n$ up to a second crossover iteration number $n_{xx}$. On the
other hand, for large $\gamma$, say for $\gamma\ge \gamma_{\tbox{C}}$, $\left< I^2
\right>$ decreases as a function of $n$ up to $n_{xx}$. Finally, in both
cases, once $n>n_{xx}$, $\left< I^2 \right>$ saturates and gets a constant
value that we call $I^2_{\tbox{sat}}$.
Moreover, we have found that $\gamma_{\tbox{C}}\approx 10^{-4}$.

In the following, the crossover iteration numbers $n_x$ and $n_{xx}$ are 
labeled with $(2)$ or $(3)$ when they correspond to $\gamma\ge \gamma_{\tbox{C}}$ 
or $\gamma<\gamma_{\tbox{C}}$, respectively.

Since we have previously concluded that Eq.~(\ref{In}), 
$I_n \approx I_0 \exp(-\gamma n)$, 
describes well the behavior of $\left\langle I \right\rangle$ vs.~$n$ for
large $\gamma$, we should try to use this fact to describe the behavior
of $\left< I^2 \right>$ vs.~$n$ when $\gamma\ge \gamma_{\tbox{C}}$; i.e., when
$\left< I^2 \right>$ decreases in the interval $n_x^{(2)}<n<n_{xx}^{(2)}$.
To this end we substitute Eq.~(\ref{In}) into Eq.~(\ref{I2a}) to write
\begin{equation}
\left< I^2 \right> 
\approx \frac{1}{n} \int_0^n I_x^2 dx =
\frac{I_0^2}{2\gamma n} [1-\exp(-2\gamma n)] \ . 
\label{I2}
\end{equation}
In fact, Eq.~(\ref{I2}) reproduces very well the numerical data for 
$\left< I^2 \right>$ vs.~$n$ when $\gamma\ge \gamma_{\tbox{C}}$ and $n<n_{xx}^{(2)}$, 
as demonstrated in Fig.~\ref{Fig7}(a) with
the blue dashed lines (we only plotted Eq.~(\ref{I2}) for the set of curves
corresponding to $I_0=10^5$ to avoid the saturation of the figure).
Moreover, since for $n>n_{xx}^{(2)}$, $\left< I^2 \right>$ saturates, we propose 
the following correction to Eq.~(\ref{I2}):
\begin{equation}
\left< I^2 \right> \approx 
\frac{I_0^2}{2\gamma n} [1-\exp(-2\gamma n)] + I^2_{\tbox{sat}} \ ,
\label{I2Isat}
\end{equation}
Also, in Fig.~\ref{Fig7}(a) we plot Eq.~(\ref{I2Isat}) in red dashed 
lines now for the set of curves corresponding to $I_0=10^4$ observing an 
excellent correspondence with the numerical data.

We want to stress that from Eq.~(\ref{I2Isat}) we can extract the dependencies
of $\left< I^2 \right>$ in the regimes bounded by $n_x^{(2)}$ and 
$n_{xx}^{(2)}$: When $n<n_x^{(2)}$, since both $\gamma$ and $n$ are small,
$\left< I^2 \right> \approx (I_0^2/2\gamma n) [1-(1-2\gamma n)]=I_0^2$;
when $n_x^{(2)}<n<n_{xx}^{(2)}$, 
$\left< I^2 \right> \approx I_0^2/2\gamma n$; while for
$n>n_{xx}^{(2)}$, $\left< I^2 \right> \approx I^2_{\tbox{sat}}$.

In addition, from Figs.~\ref{Fig7}(b-c) we conclude that 
\begin{equation}
I_{\tbox{sat}}^2\propto K^{\alpha_1} \gamma^{\alpha_2} \ ,
\label{scalingIsat}
\end{equation}
with $\alpha_1=2$ and $\alpha_2=-1$, which are in fact the same scaling exponents we 
found for $I^2_{\tbox{sat}}$ in the case 
$I_0\ll K$. It is interesting to mention that the scalings given in
Eq.~(\ref{scalingIsat}) do not depend whether $\gamma\ge \gamma_{\tbox{C}}$ or
$\gamma<\gamma_{\tbox{C}}$.

We can also estimate $n_x^{(2)}$ $\left[ n_{xx}^{(2)} \right]$ by noticing that
at $n_x^{(2)}$ $\left[ n_{xx}^{(2)} \right]$ the curves 
$\left< I^2 \right> \approx I_0^2$ 
$\left[ \left< I^2 \right> \approx I_0^2/2\gamma n \right]$ and
$\left< I^2 \right> \approx I_0^2/2\gamma n$ 
$\left[ \left< I^2 \right> \approx I^2_{\tbox{sat}} \right]$ should 
coincide. This gives 
\[
n_x^{(2)} = (2\gamma)^{-1} \ ,
\]
and
\[
n_{xx}^{(2)} = \frac{I_0^2}{2I^2_{\tbox{sat}}\gamma} \propto K^{-2} \ .
\]
 
Concerning the case $\gamma<\gamma_{\tbox{C}}$, we observe, by the use of power-law
fittings in the interval $n_x^{(3)}<n<n_{xx}^{(3)}$, that 
$\left< I^2 \right> \propto n$ (not shown here). This, together 
with $\left< I^2 \right> \propto K^2$, see the first line in Eq.~(\ref{I2K2}), gives
\[
\left< I^2 \right> \propto n K^2
\]
for $n_x^{(3)}<n<n_{xx}^{(3)}$. By matching this dependency with
$\left< I^2 \right> \approx I_0^2$ for $n<n_x^{(3)}$ and
$\left< I^2 \right> \approx I^2_{\tbox{sat}}$ when $n>n_{xx}^{(3)}$,
we estimate that
\[
n_x^{(3)} \propto I_0^2 K^{-2}
\]
and
\[
n_{xx}^{(3)} \propto \frac{I^2_{\tbox{sat}}}{K^2} \propto \gamma^{-1} \ .
\]
Finally, let us recall that the increase of $\left< I^2 \right>$
in the interval $n_x^{(3)}<n<n_{xx}^{(3)}$ is related to the non-negligible
oscillations of $I$, around its mean, when $\gamma$ is small (see Fig.~\ref{Fig6}).

\begin{figure}
\begin{center}
\includegraphics[width=8cm]{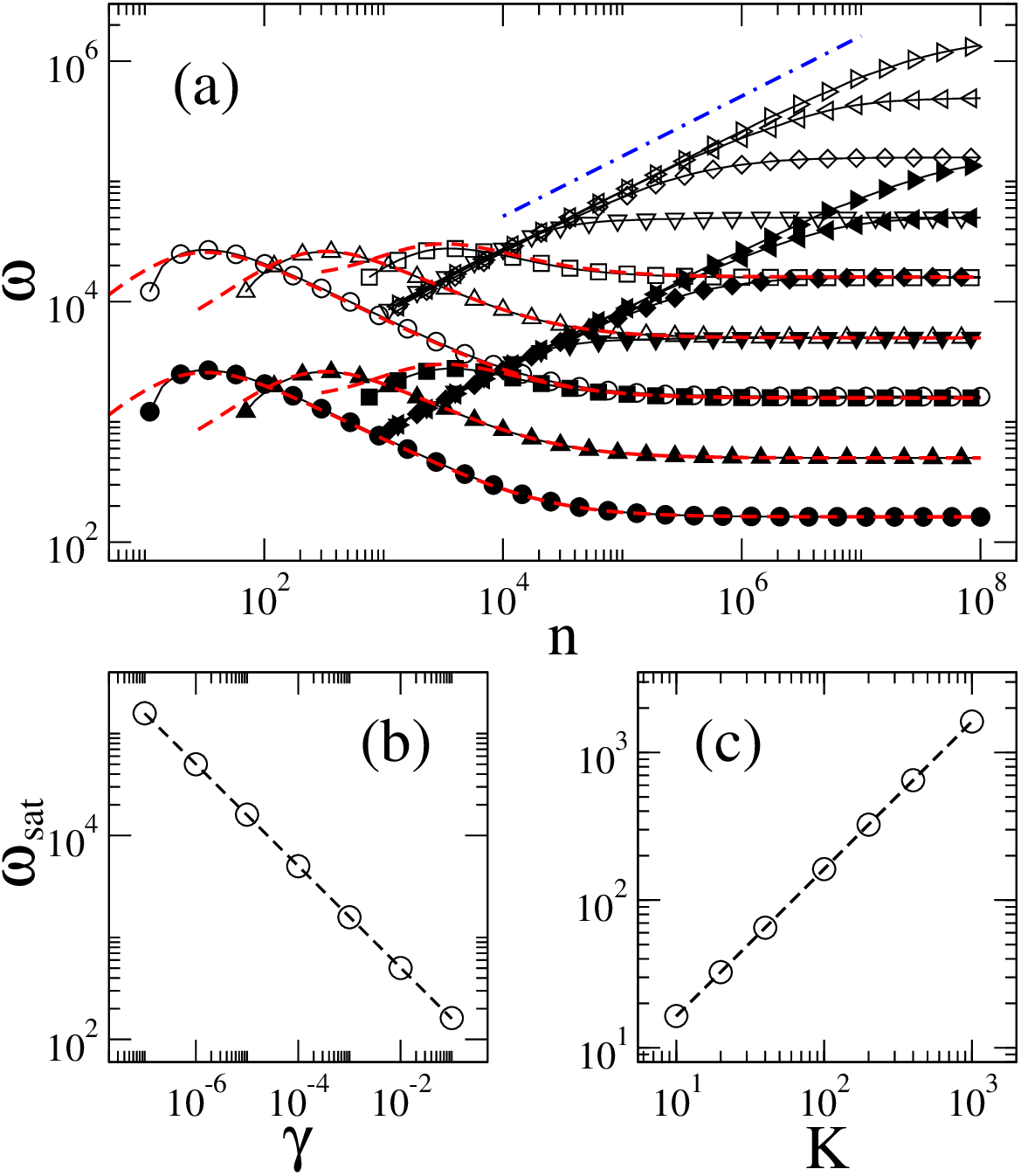}
\end{center}
\caption{(a) $\omega$ as a function of $n$ for the map of 
Eq.~(\ref{DDM}) when $K\gg1$ and $I_0\gg K$. Same parameters as in 
Fig.~\ref{Fig7}(a). The blue dot-dashed line proportional to $\sqrt{n}$ 
is plotted to guide the eye. Red dashed lines are Eq.~(\ref{w2}). (b) [(c)] 
$\omega_{\tbox{sat}}$ as a function of $\gamma$ [$K$] for $K=10^2$ and 
$I_0=10^4$ [$\gamma=10^{-1}$ and $I_0=10^4$]. The power-law fittings in (b-c),
see dashed lines, 
demonstrate the scalings $I^2_{\tbox{sat}}\propto\gamma^{\alpha_2/2}$ and 
$I^2_{\tbox{sat}}\propto K^{\alpha_1/2}$ with $\alpha_2=-1$ and $\alpha_1=2$, 
respectively.}
\label{Fig8}
\end{figure}

To complete our analysis, in Fig.~\ref{Fig8}(a) we plot $\omega$ as a function 
of $n$ for the same parameter values used in Fig.~\ref{Fig7}(a). In correspondence 
with $\left< I^2 \right>$, we found two different behaviors for $\omega$:
(i) for large $\gamma$, $\gamma>\gamma_{\tbox{C}}$, $\omega$ shows a fast increase 
as a function of $n$, reaches a maximum, and then decreases to approach a 
saturation value $\omega_{\tbox{sat}}$;
(ii) for small $\gamma$, $\gamma<\gamma_{\tbox{C}}$, $\omega$ increases 
proportional to $\sqrt{n}$ up to a crossover iteration number $n_x^\omega$, then 
for $n>n_x^\omega$, $\omega$ becomes the constant $\omega_{\tbox{sat}}$.

Since for $\gamma>\gamma_{\tbox{C}}$ we have an expression to describe $\left< I^2 \right>$
vs.~$n$ [see Eq.~(\ref{I2Isat})] and $\omega^2\equiv \left< I^2 \right>-\left< I \right>^2$
with $\left< I_n \right> = (n+1)^{-1} \sum^n_{i=0} I_i$ [see Eq.~(\ref{w})],
we can write
\begin{equation}
\omega^2 \approx 
\frac{I_0^2}{2\gamma n} [1-\exp(-2\gamma n)] 
- \left( \frac{I_0}{\gamma n} \right)^2 [1-\exp(-\gamma n)]^2 + I^2_{\tbox{sat}} \ ,
\label{w2}
\end{equation}
where we have used Eq.~(\ref{In}) to get
\[
\left\langle I \right\rangle 
\approx \frac{1}{n} \int_0^n I_x dx =
\frac{I_0}{\gamma n} [1-\exp(-\gamma n)] \ . 
\]
Then, in Fig.~\ref{Fig8}(a) we plot Eq.~(\ref{w2}), as red dashed lines, for 
$\gamma>\gamma_{\tbox{C}}$ and observe
very good correspondence with the numerical data.

\begin{figure}
\begin{center}
\includegraphics[width=8cm]{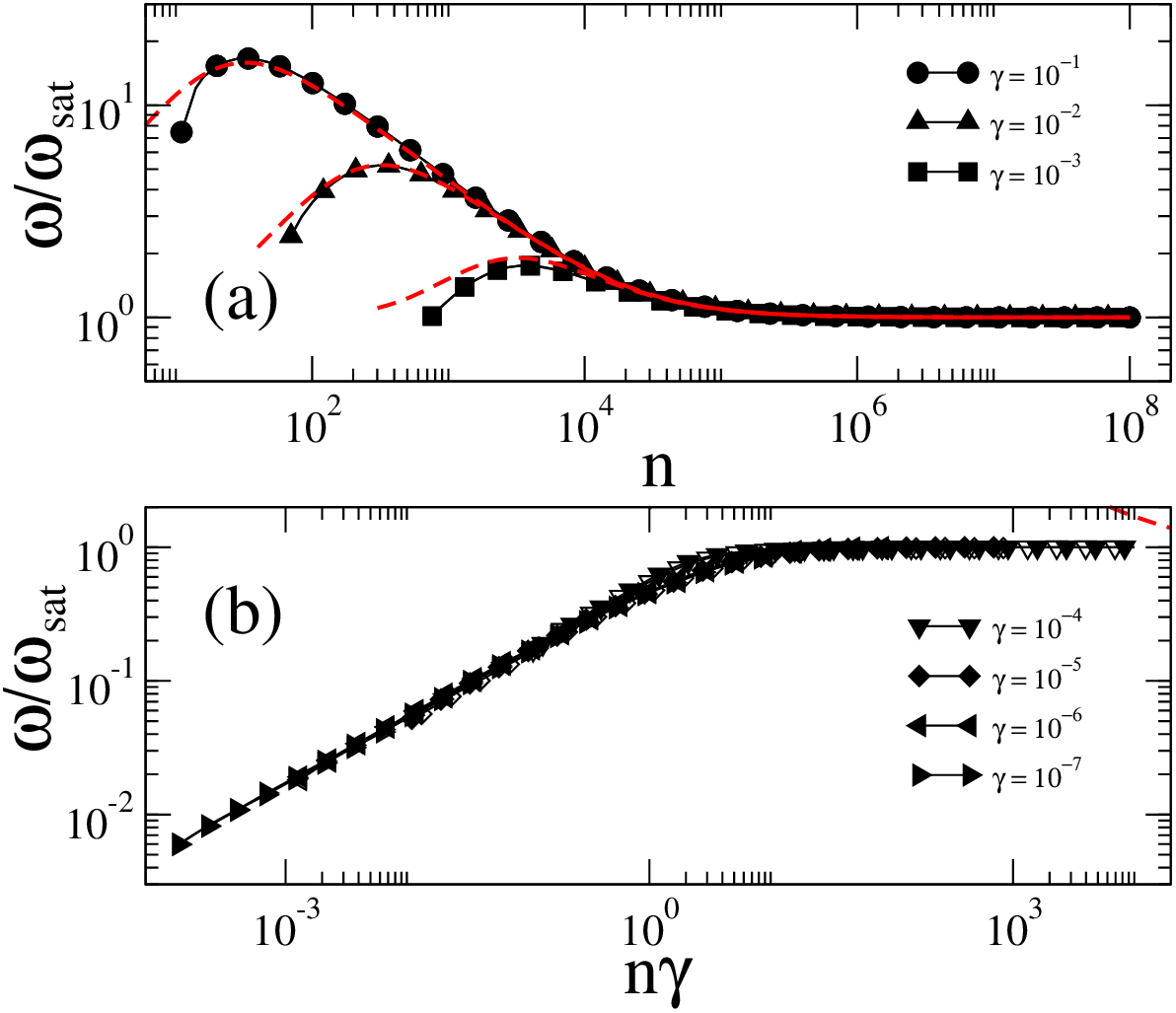}
\end{center}
\caption{(a) $\omega/\omega_{\tbox{sat}}$ as a function of 
(a) $n$ when $\gamma>\gamma_{\tbox{C}}$ and (b) $n\gamma$ when 
$\gamma<\gamma_{\tbox{C}}$ for the map of Eq.~(\ref{DDM}) when $K\gg1$ and
$I_0\gg K$. Same data as in Fig.~\ref{Fig8}(a). Red dashed lines are
Eq.~(\ref{w2}). $\gamma_{\tbox{C}}\approx 10^{-4}$}
\label{Fig9}
\end{figure}

From Eq.~(\ref{w2}) we can also see, for $n>n_x^\omega$, that
\[
\omega_{\tbox{sat}} \approx \sqrt{I^2_{\tbox{sat}}} \propto K^{\alpha_1/2} \gamma^{\alpha_2/2} \ ; 
\]
compare with Eq.~(\ref{scalingIsat}). This scaling of $\omega_{\tbox{sat}}$ is verified
in Figs.~\ref{Fig8}(b-c). Also, we have concluded that 
\[
n_x^\omega \approx n_{xx}^{(3)} \propto \gamma^{-1}
\]
(not shown). Then, we observe the collapse of the curves 
$\omega/\omega_{\tbox{sat}}$ when plotted as a function of 
(i) $n$, for $\gamma>\gamma_{\tbox{C}}$; and (ii) $n\gamma$, for $\gamma<\gamma_{\tbox{C}}$. 
See Fig.~\ref{Fig9}.

\section{Conclusions}

In this paper we studied the effects of dissipation on the scaling properties
of nonlinear discontinuous maps by analyzing the average value of the squared
action $\left< I^2 \right>$ and the deviation around the average action
$\omega$ as a function of $n$ ($n$ being the $n$th iteration of the map).
To this end we incorporate dissipation, parametrized by $\gamma$, into a 
two-dimensional discontinuous map whose dynamics is characterized by the 
stocasticity parameter $K$; see Eq.~(\ref{DDM}).
We divide our investigation in two parts, both considering the case of 
$K\gg 1$: The first is devoted to a small initial action, $I_0\cong 0$, while 
the second takes into account $I_0\gg 1$.

For $I_0\cong 0$, we described the behavior of $\left< I^2 \right>$ using 
a scaling function and analytically obtain critical exponents which are used to
overlap different curves of $\left< I^2 \right>$ and $w$ onto universal plots;
see Fig.~\ref{Fig4}. For $I_0\gg K$, we found that dissipation produces the
exponential decay of the average $I$ as a function of $n$. This fact, allowed
us to write analytical expressions describing $\left< I^2 \right>$ and $w$;
see Eqs.~(\ref{I2Isat}) and (\ref{w2}), respectively. Our results are
summarized in Table~\ref{Table1}.

\begin{table}
\begin{center}
  \begin{tabular}{|l|c|c|c|c|} \hline \hline
    & $I_0\ll K$ & $I_0\gg K$           & $I_0\gg K$ \\ 
    &  \         & $\gamma \ge \gamma_{\tbox{C}}$ & $\gamma < \gamma_{\tbox{C}}$ \\ \hline \hline
$\left< I^2 \right>\approx I^2_0$ & --- & $n<n_x^{(2)}$ & $n<n_x^{(3)}$ \\
$\left< I^2 \right>\propto nK^2$  & $n<n_x^{(1)}$ & --- & $n_x^{(3)}<n<n_{xx}^{(3)}$ \\
$\left< I^2 \right> = {1\over2} I^2_0n^{-1}\gamma^{-1}$ & --- & $n_x^{(2)}<n<n_{xx}^{(2)}$ & --- \\
$\left< I^2 \right>\approx I^2_{\tbox{sat}}$ & $n>n_x^{(1)}$ & $n>n_{xx}^{(2)}$ & $n>n_{xx}^{(3)}$ \\ \hline \hline
  \end{tabular}
\end{center}
\caption{Behavior of $\left< I^2 \right>$ for the dissipative discontinuous 
map of Eq.~(\ref{DDM}) when $K\gg 1$. We have found that
$I^2_{\tbox{sat}} \propto K^2\gamma^{-1}$, 
$n_x^{(1)} \propto \gamma^{-1}$, 
$n_x^{(2)} = {1\over2} \gamma^{-1}$,
$n_x^{(3)} \propto I_0^2 K^{-2}$,
$n_{xx}^{(2)} \propto K^{-2}$, 
$n_{xx}^{(3)} \propto \gamma^{-1}$, and 
$\gamma_{\tbox{C}}\approx 10^{-4}$.}
\label{Table1}
\end{table}

Finally, we want to add that due to the similar behavior of Chirikov's
standard map (CSM) in the regime of strong nonlinearity $K>K_{\tbox{C}}\sim 1$
and of the discontinuous map of Eq.~(\ref{DM}) in the quasilinear diffusion
regime $K>K_{\tbox{C}}=1/\tau$ \cite{MA12}, we expect the results reported
here for dissipative discontinuous maps to be also applicable to the
dissipative version of CSM with $K\gg K_{\tbox{C}}$.

{\bf Acknowledgments}.
RAS and JAMB acknowledge support form VIEP-BUAP grant MEBJ-EXC13-I  
and PIFCA grants BUAP-CA-40 and BUAP-CA-169. 
EDL thanks support from FAPESP, CNPq and FUNDUNESP, Brazilian agency.


\begin{thebibliography}{00}

\bibitem{licht} 
A. J. Lichtenberg and M. A. Lieberman,
{\it Regular and Chaotic Dynamics} 
(Springer-Verlag, New York, 1992).

\bibitem{C69}
B. V. Chirikov,
Preprint 267, Institute of Nuclear Physics, Novosibirsk (1969)
[Engl. Trans., CERN Trans. 71-40 (1971)]. 

\bibitem{C79}
B. V. Chirikov, 
Phys. Rep. {\bf 52}, 263 (1979).

\bibitem{G79}
J. M. Greene, 
J. Math. Phys. {\bf 20}, 1183 (1979).

\bibitem{M83}
R. S. MacKay, 
Physica D {\bf 7}, 283 (1983).

\bibitem{MMP84}
R. S. MacKay, J. D. Meiss, and I. C. Percival, 
Physica D {\bf 13}, 55 (1984).

\bibitem{MP85}
R. S. MacKay and I. C. Percival, 
Comm. Math. Phys. {\bf 94}, 469 (1985).

\bibitem{LS07}
D. G. Ladeira and J. K. L. Silva, 
J. Phys. A: Math. Theor. {\bf 40}, 11467 (2007).

\bibitem{ulam1}
E. D. Leonel, P. V. E. Mcclintock, and J. K. daSilva,
Phys. Rev. Lett. {\bf 93}, 14101 (2004).

\bibitem{ulam3}
D. G. Ladeira and J. K. L. da Silva, 
Phys. Rev. E {\bf 73}, 026201 (2006).

\bibitem{osiel1}
O. F. A. Bonfim,  
Phys. Rev. E {\bf 79}, 056212 (2009).

\bibitem{osiel2}
O. F. A. Bonfim,  
Int. J. Bif. Chaos {\bf 22}, 1250140 (2012).

\bibitem{well1}
E. D. Leonel and P. V. E. Mcclintock,
J. Phys. A: Math. Gen. {\bf 37}, 8949 (2004).

\bibitem{waveguide}
E. D. Leonel,
Phys. Rev. Lett. {\bf 98}, 114102 (2007).

\bibitem{other1}
D. G. Ladeira and J. K. L. da Silva, 
J. Phys. A: Math. Theor. {\bf 41}, 365101 (2008).

\bibitem{other2}
D. F. M. Oliveira and M. Robnik, 
Phys. Rev. E {\bf 83}, 026202 (2011).

\bibitem{B98}
F. Borgonovi, 
Phys. Rev. Lett. {\bf 80}, 4653 (1998).

\bibitem{stadium1}
F. Borgonovi, G. Casati, and B. Li, 
Phys. Rev. Lett. {\bf 77}, 4744 (1996).

\bibitem{stadium2}
G. Casati and T. Prosen, 
Phys. Rev. E {\bf 59}, R2516 (1999).

\bibitem{poly1}
G. Casati and T. Prosen, 
Phys. Rev. Lett. {\bf 85}, 4261 (2000).

\bibitem{poly2}
T. Prosen and M. Znidaric, 
Phys. Rev. Lett. {\bf 87}, 114101 (2001).

\bibitem{MA12}
J. A. Mendez-Bermudez, R. Aguilar-Sanchez, 
Phys. Rev. E {\bf 85}, 056212 (2012).

\end{thebibliography}
\end{document}